\documentclass[twocolumn]{aastex63}

\usepackage{ulem}
\usepackage{amsmath}
\usepackage{verbatim}
\usepackage{upgreek}

\DeclareMathOperator{\sinc}{sinc}

\newcommand{\fn}[1]{\footnote{\url{#1}}}
\newcommand{\Fig}[1]{Fig.~\ref{#1}}

\graphicspath{{./}{figures/}}

\begin{document}

\title{An analysis pipeline for CHIME/FRB full-array baseband data}

\correspondingauthor{Daniele Michilli}
\email{danielemichilli@gmail.com}

\author[0000-0002-2551-7554]{D. Michilli}
\affiliation{Department of Physics, McGill University, 3600 rue University, Montr\'eal, QC H3A 2T8, Canada}
\affiliation{McGill Space Institute, McGill University, 3550 rue University, Montr\'eal, QC H3A 2A7, Canada}

\author[0000-0002-4279-6946]{K. W. Masui}
\affiliation{MIT Kavli Institute for Astrophysics and Space Research, Massachusetts Institute of Technology, 77 Massachusetts Ave, Cambridge, MA 02139, USA}
\affiliation{Department of Physics, Massachusetts Institute of Technology, 77 Massachusetts Ave, Cambridge, MA 02139, USA}

\author[0000-0001-7348-6900]{R. Mckinven}
\affiliation{David A. Dunlap Department of Astronomy \& Astrophysics, University of Toronto, 50 St. George Street, Toronto, ON M5S 3H4, Canada}
\affiliation{Dunlap Institute for Astronomy \& Astrophysics, University of Toronto, 50 St. George Street, Toronto, ON M5S 3H4, Canada}

\author[0000-0003-2319-9676]{D. Cubranic}
\affiliation{Dept. of Physics and Astronomy, University of British Columbia, 6224 Agricultural Road, Vancouver, B.C. V6T 1Z1 Canada}
    
\author[0000-0001-6630-7871]{M. Bruneault}
\affiliation{Department of Physics, McGill University, 3600 rue University, Montr\'eal, QC H3A 2T8, Canada}
\affiliation{McGill Space Institute, McGill University, 3550 rue University, Montr\'eal, QC H3A 2A7, Canada}

\author[0000-0002-1800-8233]{C. Brar}
\affiliation{Department of Physics, McGill University, 3600 rue University, Montr\'eal, QC H3A 2T8, Canada}
\affiliation{McGill Space Institute, McGill University, 3550 rue University, Montr\'eal, QC H3A 2A7, Canada}

\author{C. Patel}
\affiliation{Dunlap Institute for Astronomy \& Astrophysics, University of Toronto, 50 St. George Street, Toronto, ON M5S 3H4, Canada}
\affiliation{Department of Physics, McGill University, 3600 rue University, Montr\'eal, QC H3A 2T8, Canada}
\affiliation{McGill Space Institute, McGill University, 3550 rue University, Montr\'eal, QC H3A 2A7, Canada}

\author[0000-0001-8537-9299]{P. J. Boyle}
\affiliation{Department of Physics, McGill University, 3600 rue University, Montr\'eal, QC H3A 2T8, Canada}

\author[0000-0001-9784-8670]{I. H. Stairs}
\affiliation{Dept. of Physics and Astronomy, University of British Columbia, 6224 Agricultural Road, Vancouver, B.C. V6T 1Z1 Canada}

\author[0000-0003-3463-7918]{A. Renard}
\affiliation{Dunlap Institute for Astronomy \& Astrophysics, University of Toronto, 50 St. George Street, Toronto, ON M5S 3H4, Canada}

\author[0000-0003-3772-2798]{K. Bandura}
\affiliation{CSEE, West Virginia University, Morgantown, WV 26505, USA}
\affiliation{Center for Gravitational Waves and Cosmology, West Virginia University, Morgantown, WV 26505, USA}

\author[0000-0002-4064-7883]{S. Berger}
\affiliation{Department of Physics, McGill University, 3600 rue University, Montr\'eal, QC H3A 2T8, Canada}
\affiliation{McGill Space Institute, McGill University, 3550 rue University, Montr\'eal, QC H3A 2A7, Canada}

\author[0000-0002-2349-3341]{D. Breitman}
\affiliation{Dunlap Institute for Astronomy \& Astrophysics, University of Toronto, 50 St. George Street, Toronto, ON M5S 3H4, Canada}
\affiliation{Department of Physics, University of Toronto, Toronto, Ontario M5S 1A7, Canada}

\author[0000-0003-2047-5276]{T. Cassanelli}
\affiliation{David A. Dunlap Department of Astronomy \& Astrophysics, University of Toronto, 50 St. George Street, Toronto, ON M5S 3H4, Canada}
\affiliation{Dunlap Institute for Astronomy \& Astrophysics, University of Toronto, 50 St. George Street, Toronto, ON M5S 3H4, Canada}

\author[0000-0001-7166-6422]{M. Dobbs}
\affiliation{Department of Physics, McGill University, 3600 rue University, Montr\'eal, QC H3A 2T8, Canada}
\affiliation{McGill Space Institute, McGill University, 3550 rue University, Montr\'eal, QC H3A 2A7, Canada}

\author[0000-0001-9345-0307]{V. M. Kaspi}
\affiliation{Department of Physics, McGill University, 3600 rue University, Montr\'eal, QC H3A 2T8, Canada}
\affiliation{McGill Space Institute, McGill University, 3550 rue University, Montr\'eal, QC H3A 2A7, Canada}

\author[0000-0002-4209-7408]{C. Leung}
\affiliation{MIT Kavli Institute for Astrophysics and Space Research, Massachusetts Institute of Technology, 77 Massachusetts Ave, Cambridge, MA 02139, USA}
\affiliation{Department of Physics, Massachusetts Institute of Technology, 77 Massachusetts Ave, Cambridge, MA 02139, USA}

\author[0000-0002-0772-9326]{J. Mena-Parra}
\affiliation{MIT Kavli Institute for Astrophysics and Space Research, Massachusetts Institute of Technology, 77 Massachusetts Ave, Cambridge, MA 02139, USA}

\author[0000-0002-4795-697X]{Z. Pleunis}
\affiliation{Department of Physics, McGill University, 3600 rue University, Montr\'eal, QC H3A 2T8, Canada}
\affiliation{McGill Space Institute, McGill University, 3550 rue University, Montr\'eal, QC H3A 2A7, Canada}

\author[0000-0001-5770-9908]{L. Russell}
\affiliation{MIT Kavli Institute for Astrophysics and Space Research, Massachusetts Institute of Technology, 77 Massachusetts Ave, Cambridge, MA 02139, USA}
\affiliation{Department of Physics, Massachusetts Institute of Technology, 77 Massachusetts Ave, Cambridge, MA 02139, USA}

\author[0000-0002-7374-7119]{P. Scholz}
\affiliation{Dunlap Institute for Astronomy \& Astrophysics, University of Toronto, 50 St. George Street, Toronto, ON M5S 3H4, Canada}

\author[0000-0003-2631-6217]{S. R. Siegel}
\affiliation{Department of Physics, McGill University, 3600 rue University, Montr\'eal, QC H3A 2T8, Canada}
\affiliation{McGill Space Institute, McGill University, 3550 rue University, Montr\'eal, QC H3A 2A7, Canada}

\author[0000-0003-2548-2926]{S. P. Tendulkar}
\affiliation{Department of Physics, McGill University, 3600 rue University, Montr\'eal, QC H3A 2T8, Canada}
\affiliation{McGill Space Institute, McGill University, 3550 rue University, Montr\'eal, QC H3A 2A7, Canada}

\author[0000-0003-4535-9378]{K. Vanderlinde}
\affiliation{Dunlap Institute for Astronomy \& Astrophysics, University of Toronto, 50 St. George Street, Toronto, ON M5S 3H4, Canada}
\affiliation{David A. Dunlap Department of Astronomy \& Astrophysics, University of Toronto, 50 St. George Street, Toronto, ON M5S 3H4, Canada}



\begin{abstract}

The Canadian Hydrogen Intensity Mapping Experiment (CHIME) has become a leading facility for detecting fast radio bursts (FRBs) through the CHIME/FRB backend. CHIME/FRB searches for fast transients in polarization-summed intensity data streams that have $24$-kHz spectral and $1$-ms temporal resolution. The intensity beams are pointed to pre-determined locations in the sky.
A triggered baseband system records the coherent electric field measured by each antenna in the CHIME array at the time of FRB detections. 
Here we describe the analysis techniques and automated pipeline developed to process these full-array baseband data recordings.
Whereas the real-time FRB detection pipeline has a localization limit of several arcminutes, offline analysis of baseband data yields source localizations with sub-arcminute precision, as characterized by using a sample of pulsars and one repeating FRB with known positions. The baseband pipeline also enables resolving temporal substructure on a micro-second scale and the study of polarization including detections of Faraday rotation.

\end{abstract}

\keywords{}


\section{Introduction} \label{sec:intro}
Fast radio bursts \citep[FRBs;][]{lor07} are millisecond-duration radio transients whose integrated column density of free electrons (approximately quantified by the dispersion measure, DM) significantly exceeds the maximum value expected by Galactic models \citep{cor02, yao17}.
The extragalactic nature of the bursts has been confirmed with the localization of a handful of FRBs to host galaxies\footnote{\url{http://frbhosts.org} (Heintz et al., in prep.)}.
The cosmological distance of these host galaxies gives severe constraints on source emission models, requiring radio luminosities far higher than those of any known Galactic source.
A comprehensive explanation of the phenomenon is still missing and a variety of models exist\footnote{\url{https://frbtheorycat.org} \citep{pla18}}.

The Canadian Hydrogen Intensity Mapping Experiment \citep[CHIME; ][]{new14} radio telescope has been equipped with the CHIME/FRB backend \citep{chi18_overview} to discover and study FRBs.
Thanks to the large instantaneous field of view ($\sim$ 200 square degrees), high sensitivity and almost continual on-sky time, CHIME/FRB has already detected hundreds of FRBs \citep[e.g.][]{fon20}.
CHIME/FRB blindly searches for sources with a real-time pipeline capable of processing the data from the telescope and automatically selecting potential candidates \citep{chi18_overview}.
The data processed by the real-time pipeline has a resolution of $\sim 1$\,ms and no polarization information. 
However, extracting as much information as possible for every burst is essential since the vast majority of FRB sources are detected just once\footnote{\url{http://frbcat.org} \citep{pet16_frbcat}} and only a few have been reported to repeat \citep{spi16,chi19_r2,chi19_8repeaters,kum19,fon20}.

It is interesting to study FRBs at very high resolution for understanding, among other things, the complex and puzzling time-frequency structures of their signal \citep{hes18}, and scintillation of the signal by intervening plasma \citep{mas15}.
Also, polarization information is key to understanding the emission mechanism and local environment of FRBs \citep{pet15, mas15}.
Finally, source localization with the static FFT beams of the real-time search \citep{chi18_overview,ng17} has uncertainties of roughly the beam width of tens of arcminutes \citep[e.g.][]{chi19_8repeaters}.
However, many of the scientific goals of CHIME/FRB rely on an accurate localization of the detected FRBs. 
Some examples are the cross-correlation of FRBs with structure in the Universe \citep{mas15b, raf20}, the study of the luminosity function of FRBs, where a localization is essential to compensate for beam effects when measuring the flux of the detections, and the localization of bright, low-DM FRBs to host galaxies in the local Universe. 

To allow detailed studies of FRBs, a triggered baseband recording system capable of storing the electric field measured by each of the 1024 dual-polarization feeds has been developed for CHIME/FRB \citep{chi18_overview}.
Here, we report the operations of the baseband system of CHIME/FRB and describe the analysis techniques and software developed to process these data.
A summary of the baseband system is provided in \textsection\ref{sec:capturing_system}.
Algorithms to form tied-array beams and correct for signal dispersion are described in \textsection\ref{sec:beamforming}.
The process to use these beams to refine source positions is detailed in \textsection\ref{sec:localization}.
An automated pipeline that processes the baseband data by using the aforementioned software is outlined in \textsection\ref{sec:pipeline}.
The localization capability of the baseband pipeline is characterized in \textsection\ref{sec:characterization}.
Finally, conclusions are drawn in \textsection\ref{sec:conclusions}.

\section{CHIME/FRB Triggered Baseband Recording System}
\label{sec:capturing_system}
The detection pipeline of CHIME/FRB searches for short-duration, dispersed peaks in a stream of total intensity data in each of 1024 formed beams \citep{ng17,chi18_overview}.
\citet{chi18_overview} described the plan to create a triggered baseband recording system for the CHIME/FRB experiment. 
The system is now in place and routinely records snapshots of baseband data around signals of interest.
The raw voltages measured by the 1024 dual-polarization feeds of the telescope, which operates between 400 and 800\,MHz, are amplified, digitized and processed by a cluster of motherboards called the F-engine \citep{ban16_fengine}. 
For each receiver, the F-engine uses a 4-tap polyphase filter bank to produce a spectrum with 1024 channels (each 390 kHz wide) every $2.56$\,$\upmu$s.
The baseband data are calculated by rounding this output to 4+4 bit complex numbers. 
The resulting data rate is 6.5 Tb/s and a memory buffer allows the storage of $35.5$ seconds of baseband data at a given time.
From the moment a signal arrives at the telescope, the real-time pipeline is able to process the event and trigger a baseband dump in $\sim 14$ seconds.
This leaves a usable data buffer of $\sim 20$ seconds, which corresponds to a maximum DM of $\sim 1000$\,pc\,cm$^{-3}$ at CHIME frequencies.
Typically, $100$\,ms of baseband data are stored around the time of arrival (TOA) of triggering events for each frequency channel. 
Additional time is usually added around the burst due to the uncertainties in the TOA and DM values reported by the real-time pipeline, which are between $16$--$128$\,ms and $1.6$--$25$\,pc\,cm$^{-3}$, respectively.
This produces an average of $\sim 100$\,GB of baseband data stored for a detected event.

We have modified the threshold on the S/N to trigger a baseband dump in different periods to reduce or increase the number of dumps per day.
With thresholds on the S/N ranging between 8-10, the typical number of baseband dumps per day has consequently varied between $\sim$1-10. 
There is also the possibility to insert special rules to trigger, for example, known pulsars that are otherwise discarded.
If needed, baseband dumps can also be triggered manually for study of specific sources.

\section{Beamforming and de-dispersion}\label{sec:beamforming}
The phase information contained in the baseband data can be used to phase-reference detected waves to any direction within the field of view of the telescope prior to coadding over feeds.
This process, referred to as \emph{beamforming}, maximizes the telescope sensitivity to a certain direction and effectively \emph{points} the telescope to that direction.
Within the CHIME experiment, a bright source is observed every day to calibrate the gains of each receiver of the telescope \citep{chi18_overview}.
Our baseband pipeline uses by default the closest calibration observation in time as a phase reference. 
A number of beams is then formed in a set of directions by adding the expected geometrical delays with respect to the geometrical center of the telescope.
This operation is performed by using custom developed software.
The Python ephemeris library  \texttt{Skyfield}\footnote{\url{https://rhodesmill.org/skyfield}} is used to calculate the geometrical factors for beamforming and calibration.
The beamforming routine is the most computationally expensive stage of the baseband pipeline, with one beam formed in roughly one hour with the current configuration. 
The computational time and memory needed to form the beams limit us  to $\sim 200$ beams formed simultaneously with our current processing machine. 

The dispersion of radio signals induced by intervening free electrons along the line of sight impairs the detection of fast transients and it is usually corrected for with de-dispersion algorithms.
Traditionally, this operation is performed \emph{incoherently} by adding appropriate delays to different frequency channels relative to a reference frequency \citep[e.g.][]{lor04}.
However, residual smearing of the signal within single channels will diminish the signal-to-noise ratio (S/N) of the event and smooth out temporal features shorter than the smearing time.\footnote{For a source with $\text{DM}=500$\,pc\,cm$^{-3}$ detected by the real-time pipeline of CHIME/FRB, which has 16,384 channels, the smearing time at the bottom of the band is $\sim 0.4$\,ms, while at the native resolution of baseband data, which have 1024 channels, the smearing time at the bottom of the band is $\sim 25$\,ms.}
With the use of baseband data, it is possible to mathematically remove the effect of dispersion by modelling the intervening medium as a cold tenuous plasma \citep{han71}. This leads to the following transfer function, whose inverse can be used to rotate the phase of the incoming radio waves and effectively correct for intra-channel smearing,
\begin{equation}
    H(\nu+\nu_0) = \exp\left(\frac{2\pi\textrm{i}\nu^2k_{\text{DM}}\text{DM}}{\nu_0^2(\nu+\nu_0)}\right),
\end{equation}
where $\nu_0$ is a reference frequency, $\nu$ is the frequency offset from $\nu_0$ and $k_{\text{DM}}^{-1} = 2.41\times10^{-4}$\,MHz$^{-2}$\,pc\,cm$^{-3}$\,s$^{-1}$ \citep{man72} is the dispersion constant.

\begin{figure*}
\plotone{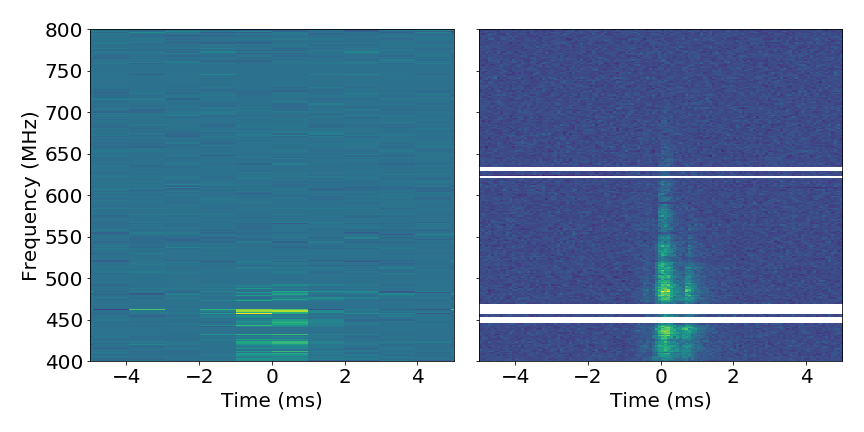}
\caption{
Example of a burst from FRB 20190618A (CHIME/FRB Collaboration, in prep.) detected by CHIME/FRB and processed through the intensity (left) and baseband (right) pipelines. 
The color scale, independent for the two plots, is proportional to the intensity of the signal.
The left plot has a time resolution of $\sim$ 1 ms and has been incoherently dedispersed to a DM of $228.93$\,pc\,cm$^{-3}$. 
The right plot has a time resolution of $81.92$\,$\upmu$s (i.e. 32 times the native time resolution) and has been coherently dedispersed to the same DM value. 
Both panels are plotted with 256 frequency channels.
Horizontal white lines in the right plot are frequency channels missing due to the failure of one of the machine's nodes during the baseband dump.
\label{fig:waterfall}}
\end{figure*}

We use baseband data to coherently correct for the dispersion of radio signals, which allows us to resolve the bursts at a high temporal resolution.
An example of a comparison between the waterfall plots -- the signal intensity displayed as a function of time and frequency -- of the same burst obtained with intensity and baseband data, respectively, is displayed in Fig.~\ref{fig:waterfall}. 
The burst shown is a CHIME/FRB detection of FRB 20190618A (CHIME/FRB Collaboration, in prep.).
The spectro-temporal structures in the burst, unresolved by the intensity data, are clearly visible in the baseband data. 
This is due to a combination of improvements made possible by the baseband data, namely the bandpass correction by phasing the telescope to the source location,\footnote{The telescope primary beam will affect baseband and static FFT beams alike. However, its size is much larger than the size of formed beams.} the higher temporal resolution and the ability to coherently correct for the signal dispersion.

\section{Localization}
\label{sec:localization}
CHIME/FRB formed beams have a $\text{FWHM} \sim 15$--$30$\,arcmin across the band at zenith.
However, due to the interferometric nature of the telescope, the localization of point sources can be improved by mapping the signal intensity around the initial detection.
This intensity map can then be fitted with a model of the expected telescope response to obtain a more accurate position.
By using this technique, it is theoretically possible to obtain a precision on the localization given by \citep{mas19}
\begin{equation}\label{eq:loc_theoretical_limit}
    \sigma_\theta = \frac{\sqrt{6}}{2\pi}\frac{\lambda}{D}\frac{1}{\text{S/N}}\frac{1}{\cos(\theta)},
\end{equation}
where $\lambda$ is the observing wavelength, $D$ is the maximum separation between the telescope antennae, S/N is the signal-to-noise ratio of the detected event and $\theta$ is the zenith angle of the source.
For a source detected at zenith by CHIME/FRB, where $D \sim 80$\,m and considering $\lambda = 600$\,MHz, this translates to 
\begin{equation}\label{eq:loc_theoretical_limit_chime}
    \sigma_\theta \sim \frac{8}{\text{S/N}}\,\text{arcmin}.
\end{equation}
Therefore, a bright FRB with a $\text{S/N}=100$ might be localized by CHIME/FRB with a precision of $\sim 5$\,arcsec.
However, unmodelled systematic effects will limit the localization accuracy, as discussed in \textsection\ref{sec:characterization}.

Our strategy to map the signal intensity is to phase-reference the detected waves to a grid of trial positions around the initial localization and to measure a single value of the S/N value in each of them.
This corresponds to forming a grid of overlapping beams in the local reference system of the telescope. 
The total S/N of the burst measured in each beam is calculated and the resulting array is then fitted with a model describing the beam response of the telescope.
Currently, we are using a 2D $\sinc^2$ function with symmetry axes aligned to the East-West (EW) and South-North (SN) directions and five free parameters (peak position, two widths, and the height of the curve).
This model has been found to fit well the sensitivity response of CHIME formed beams within their FWHM, with typical residuals $\lesssim 5\%$.
We are currently investigating the use of a more realistic beam model, obtained by simulating the expected response function of all the single telescope's antennae, and it will be presented in a future paper.
Since the noise in different beams is not independent, we use a covariance matrix in the fit across the different beams calculated by using Eq.~38 of \citet{mas19}.

\begin{figure*}
\plotone{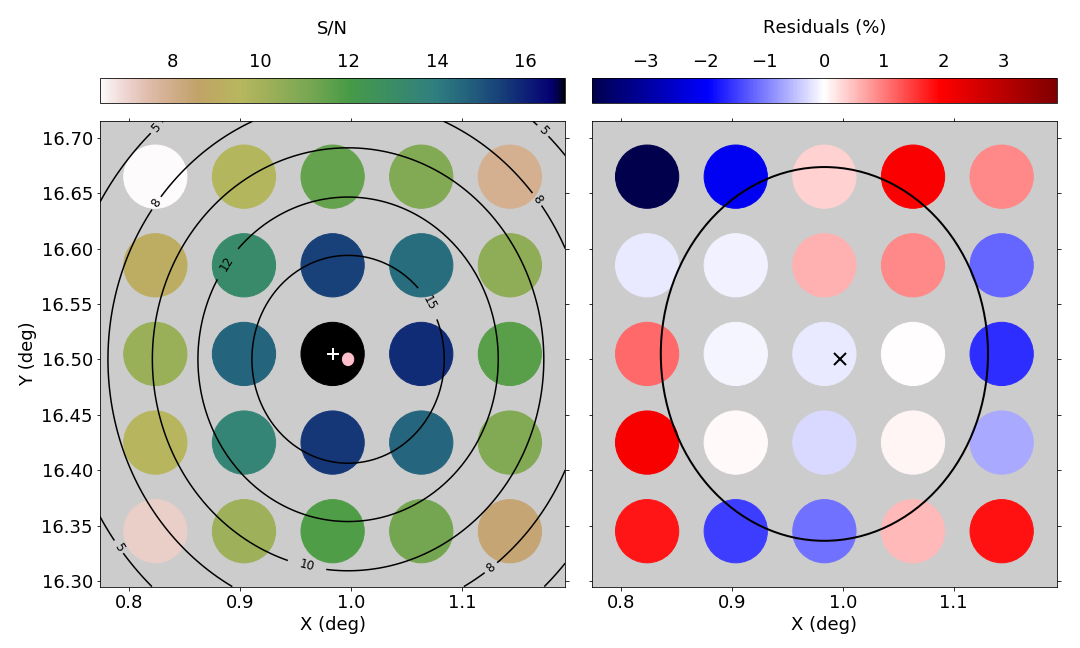}
\caption{
Localization of a relatively weak burst from FRB~20180916B \citep{chi19_8repeaters} before applying any post-localization correction, as discussed in \textsection\ref{sec:characterization}. The localization is presented in the local coordinates of the telescope, with X going from East to West and Y from South to North, referenced to $400$\,MHz.
Left: Intensity map of the signal measured with a grid of tied-array beams fitted with a Gaussian model of the telescope beam. 
Colored circles are placed at the position of the beam centers at 400\,MHz and their color-scale is proportional to the S/N detected in each beam, as indicated by the color bar at the top. 
Black contours represent the Gaussian model at different levels, reported as labels on the plot.
The central pink ellipse, with a size of $\sim 0.4$\,arcmin, is the 1$\sigma$ localization region from the fit.
The true source position measured by \citet{mar20} is located at the center of the plot and depicted with a white `+' symbol.
Right: Percentage residuals of the fit; the color-scale is reported in the color bar at the top.
The black cross represents the fit localization, while the black curve is the FWHM of the central beam at 600\,MHz.
\label{fig:localization_example}}
\end{figure*}

An example of the localization of a relatively weak single pulse from FRB 20180916B \citep{chi19_8repeaters} is shown in \Fig{fig:localization_example}.
The coordinate system used in the plot, where the grid of beams is regularly spaced, is a local spherical-polar coordinate system centered at the telescope zenith where X runs EW, perpendicular to the telescope meridian, and Y runs SN, parallel to the telescope meridian, referenced to $400$\,MHz.
The different frequency channels of single beams are aligned to have the same RA and Dec at the beam center.


\section{Baseband processing pipeline}
\label{sec:pipeline}
\begin{figure}
\plotone{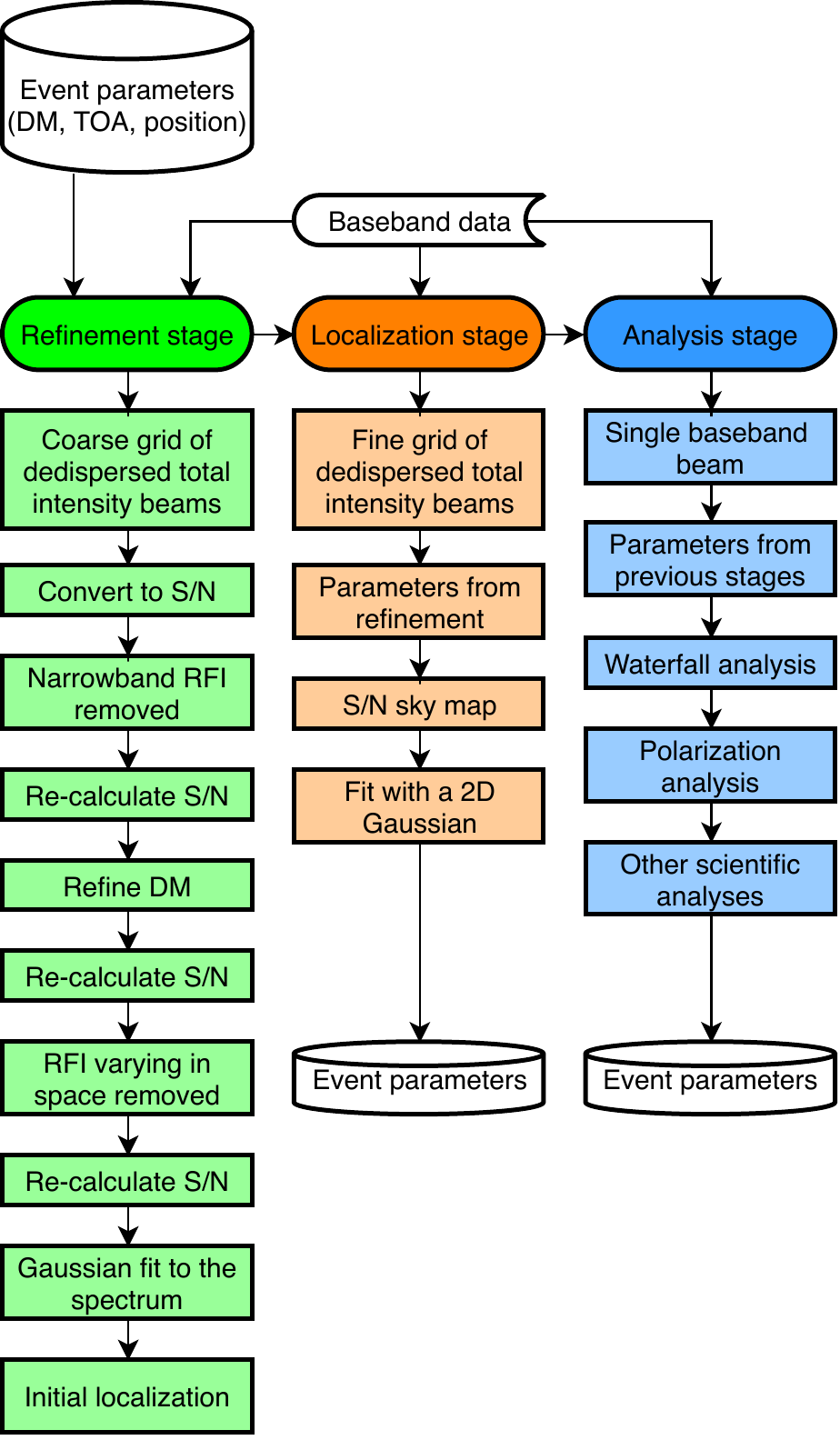}
\caption{Schematic of the automated pipeline used to process CHIME/FRB baseband data.
The colors represent the three different stages of the pipeline, indicated at the top of each branch and highlighted with darker shades and rounded symbols.
Single tasks forming the three stages are depicted below each branch and are highlighted with lighter colors and squared symbols. 
Data and metadata are depicted in white and with different symbols. See text for details.
\label{fig:pipeline_scheme}}
\end{figure}

A framework with minimal overhead was developed to pipeline the analysis described in the previous sections to produce scientifically useful output from the baseband data recorded by CHIME/FRB without any human intervention. 
The different steps involved in this pipeline are presented in \Fig{fig:pipeline_scheme} and described in detail below. The framework and pipeline are hardware agnostic and designed to be highly scalable and are currently operational on a compute cluster located at the telescope. 
The baseband pipeline can run automatically on the new events that are identified by the real-time detection pipeline of CHIME/FRB or be manually triggered to process specific events.

\subsection{Pipeline infrastructure}
The automated pipeline consists of three major technologies, Apache Airflow,\fn{https://airflow.apache.org} an open source platform to schedule and monitor workflows; Docker,\fn{https://www.docker.com} a platform to deliver software in standard, environment agnostic, packages, called containers; and a custom built coordinator software which manages metadata queuing and orchestration of docker containers through Docker Swarm\fn{https://docs.docker.com/get-started/orchestration}. 
When an event is detected by the real-time pipeline and the relevant baseband data are stored on disk, a unique identifier for the event is stored in a \emph{bucket} of the memory of the coordinator. This unique identifier, called the \emph{event ID}, is the key to retrieving and depositing all event information in a database of CHIME/FRB detections. 
The Airflow workflow, comprising of application programming interface (API) calls to the coordinator in a directed acyclic graph (DAG), scans the baseband bucket periodically (currently, every 15 minutes) and instantiates the analysis pipeline if an event ID is found. Additionally, the coordinator allows the user to manually insert a list of event IDs to be processed into this bucket.

At this point, following the recipe described in the DAG, Airflow launches a series of analysis tasks to be performed on the cluster via the coordinator API. In order to optimize hardware resource utilization while not sacrificing consistency of results, all cluster tasks are mapped to a Docker container. This provides the capability to transform and adapt the analysis pipeline around resource bottlenecks for tasks with significantly different CPU, RAM and I/O requirements. Although the pipeline is designed to be hardware agnostic, in its current configuration it processes the 1024 frequency channels of a baseband dump over 64 containers, each with an allocation of 2 cores and up to 16 GB of RAM (depending on the dump length) running in parallel on the analysis cluster, where each node consists of a dual socket Intel\textsuperscript{\tiny\textregistered} Xeon\textsuperscript{\tiny\textregistered} CPU E5-2630 v4 2.20\,GHz and 128\,GB of RAM.

In each container, two frequency channels are analyzed in parallel at a time until all the channels are processed. A separate dedicated task monitors the status of all the analysis containers and performs any ancillary post-processing routines. This configuration was chosen to reduce the memory pressure on the cluster but could be easily modified to optimize for CPU cycles or I/O bandwidth by either spawning a different number of containers or by modulating the hardware resource budget for each container. Additionally, due to the distributed design of the pipeline framework, multiprocessing designs can be avoided in favor of ease of prototyping, reduced complexity and more accessible single-thread performance.

\subsection{Pipeline tasks}
The baseband analysis pipeline is composed of three stages: refinement, localization and analysis, as shown in \Fig{fig:pipeline_scheme}, each consisting of multiple tasks, as described in the following sections.
Each stage takes approximately one hour to complete on the current analysis cluster, with the time budget being dominated by beamforming operations described in \textsection\ref{sec:beamforming}.

\subsubsection{Refinement stage}


At the refinement stage, the initial parameters of the signal measured by the real-time pipeline are improved using the baseband data. 
The uncertainties measured by the real-time pipeline are between $16$--$128$\,ms for the burst TOA, between $1.6$--$25$\,pc\,cm$^{-3}$ for its DM and $\sim 15$\,arcmin for its position.
If available, we use as initial localization guess the sky position derived by a parallel offline analysis pipeline called the \emph{intensity localization pipeline}. 
The intensity localization pipeline refines the source sky position measured by the real-time pipeline by using the frequency-resolved total intensity of the signal detected by multiple static FFT beams. This increases the position accuracy of the source to a few arcminutes (Scholz et al., in prep.).

A set of beams is formed around the initial position and coherently dedispersed at the initial DM estimate. The grid of beams, used to refine the initial localization, is currently configured to produce 3 beams in the NS direction and 10 beams in the EW direction. The larger number in the EW direction is chosen because of the smaller number of independent baselines of the telescope in this direction, which makes the initial EW guess uncertain. The total size of the grid is $\sim 1.1 \times 0.25$ deg$^2$. 
It may happen, however, that the initial position is too far away from the true source position, especially for sources detected in far side-lobes of the primary telescope beam.
In these rare cases, it is necessary to manually search for a better initial position.
We are currently working to further increase the tolerance of the refinement stage to initial positions with large offsets by using an MCMC code (Scholz et al., in prep.).

The total intensity is calculated for each beam as a function of frequency and time. The resulting array is downsampled in time such that the pulse width measured by the real-time pipeline spans three bins. 
The S/N is calculated for each frequency channel and time bin by normalizing the off-pulse RMS to zero average and unitary standard deviation independently in each frequency channel and beam. In order to find the off-pulse region, an average profile is calculated and any time bin above a S/N of three is removed.
This algorithm is applied iteratively in three steps to gradually improve the S/N.
In the first step, narrow-band radio-frequency interference (RFI) is excised.
This is done by calculating an average S/N for each channel and time bin across all the beams and then using a metric given by the standard deviation divided by the average in each channel. 
After normalizing this metric to have zero median and unity RMS, and after removing an arbitrary slope to account for intrinsic spectral variations, we remove channels above a specified threshold, currently set to three.
This process is iterated upon until no channels above the threshold are found.
In the second step, the initial DM value is incoherently refined with a brute-force algorithm that maximizes the S/N with a resolution of $0.1$\,pc\,cm$^{-3}$. 
Finally, the RFI mask is refined by using the intensity variation of the signal across multiple beams.
This is performed by calculating an average value of the RMS in each beam and channel.
For each channel, we define a metric given by the fractional difference between the maximum and minimum value of the total intensity across all the beams.
In an iterative process, we normalize the metric across all channels and mask values larger than three.
This technique is effective in removing RFI for which the intensity changes rapidly in different sky directions.
The same RFI mask is applied to all the beams.
After each of these three steps, the S/N array is re-calculated with the algorithm described above.
In order to increase the S/N of narrow-band bursts, a Gaussian function is fitted to the spectrum of the beam displaying the strongest signal. If the maximum value of the Gaussian within the bandwidth is larger than 3 times the RMS, channels outside 3 times the standard deviation of the function are removed.

In order to refine the initial source localization, a single S/N value is calculated for each beam of the grid. 
The position of the beam with the highest S/N value is used as the initial guess position for the subsequent localization stage.

\subsubsection{Localization stage}
As discussed in \textsection\ref{sec:localization}, the source position is further refined by forming a compact grid of largely overlapping beams to map the sky region around the initial guess determined in the previous stage.
We tested the optimal size and filling factor of the beam grid to optimize computing power and localization precision.
A good compromise in terms of number of beams and localization precision was found to be a grid composed of $5\times5$ beams covering a square sky region with a side of $\sim 0.3$\,deg, i.e. similar to the FWHM of a single beam, as illustrated in Fig.~\ref{fig:localization_example}.

The beams are dedispersed to the DM calculated in the refinement stage and the rest of the parameters obtained there are applied.
A single pulse profile is obtained for each beam by averaging the frequency channels together.
The time bin corresponding to the strongest S/N value across all the beams is selected.
The S/N of the profiles in different beams corresponding to the same time bin is used to form a map of the signal intensity at different locations on the sky. A 2D Gaussian function approximating the beam shape is then fitted to this S/N map to obtain the refined source position and its uncertainty, as described in \textsection\ref{sec:localization}.

\subsubsection{Analysis stage}
A single tied-array beam is formed at the source direction calculated in the localization stage.
The parameters obtained in the refinement stage are once again used to maximize the S/N.
The resulting data product is a complex-valued array resolved in time, frequency and polarization, with the native time and frequency resolutions presented in \textsection\ref{sec:capturing_system}.
To facilitate subsequent scientific analyses, this complex array is also converted to power values containing information on the four Stokes parameters.

A series of additional analyses is performed on these arrays to produce scientific results. 
The analyses include the characterization of the polarization properties of the signal, such as Faraday rotation, polarization fraction and polarization angle. 
This will be detailed in a forthcoming paper (Mckinven et al., in prep.)
The higher temporal resolution allows us to study the spectro-temporal properties of the bursts in detail, theoretically up to nanosecond timescale.
This analysis is also important to study various propagation effects on the bursts.
For example, potential strong gravitational lensing of the bursts at millisecond timescales is being studied by correlating the phases of beamformed baseband data as a function of time-lag.
The baseband data of CHIME/FRB will also be used in the future to correlate the TOA of the signal with outrigger telescopes to perform very-long-baseline interferometry (VLBI) localization of the sources with an expected precision smaller than $\sim$ 50 milli-arcseconds \citep[e.g.][]{leu20}.

At the end of the pipeline, the science output is stored in a centralized database of CHIME/FRB detections and diagnostic plots are displayed in a web viewer for inspection.

\section{Characterization of the localization capability}
\label{sec:characterization}
When localizing a source, it is essential to report a reliable uncertainty region.
In addition to statistical uncertainties, a number of systematic effects are expected to affect the localization of FRBs detected by CHIME/FRB.
Some of these systematics can be modeled and corrected while others remain poorly understood and must be estimated through sources with known positions. The methodology for estimating these systematics is presented below. 

Firstly, using interferometric observations of bright sources collected with the cosmology backend of CHIME \citep{new14}, we identified an unplanned rotation of the telescope’s structure about an axis oriented parallel to the local zenith, about 4 arcminutes clockwise as viewed looking down this axis towards the Earth.
This was corrected by updating the assumed feed positions prior to beamfoming.
Localization errors of a few arcminutes are also expected from the thermal expansion of the metal structure of the telescope in the SN direction, which uniformly scales with the baseline lengths. 
The thermal expansion coefficient was similarly measured using interferometric data collected over a range of ambient temperatures. 
We correct single-pulse localization for this effect post-facto by using the measured expansion coefficient and the ambient temperature at the time of observation.
A small pointing error, up to a few tenths of an arcminute, is also expected from time-variable $\sim 10$\,ps errors in the distribution of the telescope clock between the digitizers for feeds on the east two cylinders (located in the East Receiver Hut) and the west two cylinders (located in the West Receiver Hut). These clock errors are continuously monitored by injecting a common noise signal into a digitizer located in each hut and examining the phase of the cross-correlation. Using these monitoring data, the induced localization error is also corrected post-facto.

In addition to the effects we have accounted for, there are several other effects that could degrade our localization accuracy. 
For example, we have calibrated the telescope orientation around the vertical axis, but have not accounted for any tilt in either the NS or EW directions. 
Nonuniformities in the telescope structure could also cause offsets in the localization.
Also, feed-to-feed variations in the primary beam will introduce a direction-dependent phase that is difficult to correct and will limit localization precision. These beam variations are currently being mapped through a campaign of holographic interferometry \citep{new14,ber16}, required for CHIME's cosmology science. These measurements could be used to assess and even correct phase variations prior to beamforming. This remains to be pursued and currently no explicit attempt is made to correct for this effect.
The consequence of these and all other unmodelled systematics is characterized by localizing pulses from multiple pulsars and the repeating FRB 20180916B, whose positions were accurately determined by other instruments.

\begin{figure*}
\plotone{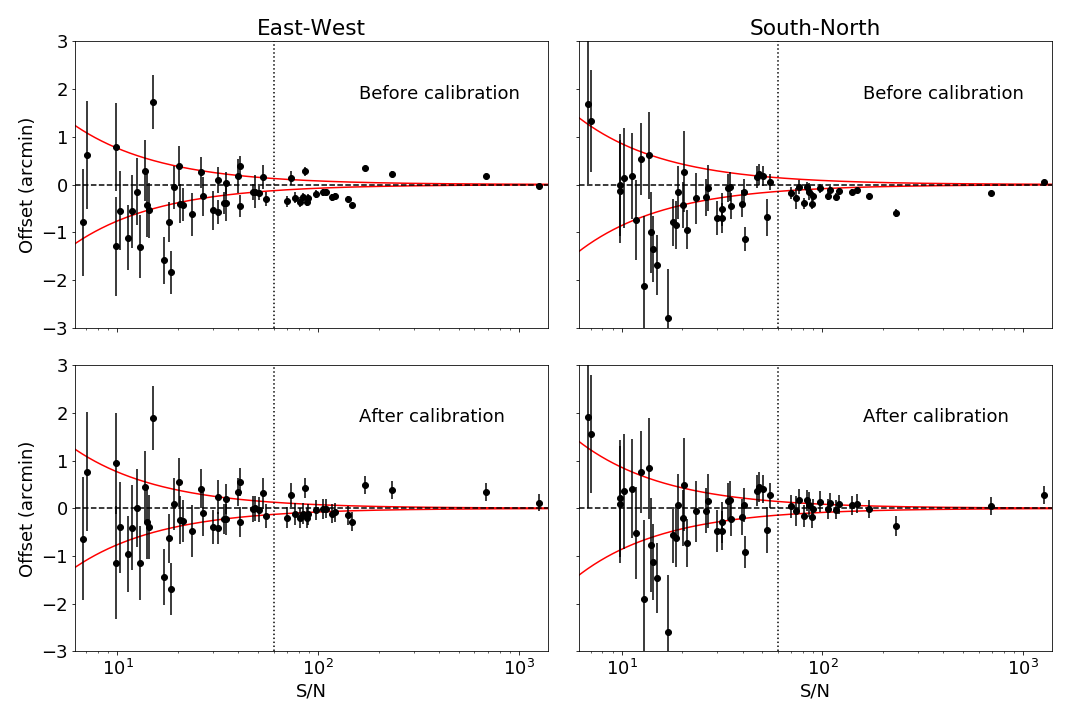}
\caption{
Measured offset of pulses from the sources presented in Table~\ref{tab:pulsar_sample} with respect to their expected position as a function of the detected S/N.
The offset is presented in the local reference system of the CHIME telescope. Left: offset in the EW direction; right: offset in the SN direction. 
The top panels report values and uncertainties obtained after including all modelled systematic effects, while the localizations in the bottom panels have been calibrated to have zero weighted average and unitary $\chi^2_\text{red}$.
The horizontal dashed line highlights zero offsets, the vertical dotted line represents an empirical cutoff between weak and bright pulses.
Red curves are the theoretical uncertainty limits obtained from Eq.~\eqref{eq:loc_theoretical_limit_chime} for sources at zenith. Note that some of the pulses are detected away from zenith and, therefore, their theoretical uncertainties are larger by a factor $\cos(\theta)^{-1}$ (Eq.~\ref{eq:loc_theoretical_limit}).
\label{fig:loc_uncert}
}
\end{figure*}

After applying all modelled localization corrections, we use a sample of 56 pulses from 15 sources with a known position to perform an empirical calibration of our localizations and their uncertainty. 
This sample is presented in Table~\ref{tab:pulsar_sample}.
The sources have been chosen to cover a large fraction of CHIME parameter space in terms of S/N and declination.
We plan to further expand this sample in the future to have an even better coverage of the telescope parameter space.
We assume that our telescope coordinates X and Y (described above) are the natural coordinates that describe the space in which systematic errors live.
The empirical calibration transforms our localizations as follows:
\begin{align} \label{eq:loc_calibration}
    \theta^i_x &\pm \sigma^i_x \rightarrow \left(\theta^i_x + \delta_x\right) \pm \left(\sqrt{\left(\alpha \sigma^i_x\right)^2 + \Sigma_x^2}\right)
    \nonumber \\
    \theta^i_y &\pm \sigma^i_y \rightarrow \left(\theta^i_y + \delta_y\right) \pm \left(\sqrt{\left(\alpha \sigma^i_y\right)^2 + \Sigma_y^2}\right),
\end{align}
where the index $i$ runs over the pulses in the sample, and the calibration has 5 free parameters: a scaling of the statistical uncertainty $\alpha$, systematic offsets in each direction $\delta_x$ and $\delta_y$; and a systematic floor in the uncertainty in each direction $\Sigma_x$ and $\Sigma_y$. 
The parameter $\alpha$ is expected to be $\sim 1$ and eventual deviations could arise e.g. from imperfect modeling of the data, such as non-Gaussianities in the noise or deviations of the beam response with respect to the mathematical model.
We adjust the parameters in Eq.~\eqref{eq:loc_calibration} by dividing our sample in weak and bright bursts and following the criteria:
\begin{enumerate}
    \item The weighted mean of the position errors is zero (affects mainly $\delta_x$ and $\delta_y$).
    \item The reduced $\chi^2$ is unity for low S/N events (affects mainly $\alpha$).
    \item The reduced $\chi^2$ is unity for high S/N events (affects mainly $\Sigma_x$ and $\Sigma_y$).
\end{enumerate}
Therefore, we set four conditions (i.e. unitary $\chi^2_\text{red}$ on weak and bright bursts for the two independent coordinates), fit for the three parameters $\alpha$, $\Sigma_x$ and $\Sigma_y$, and imposing the weighted average on X and Y coordinates to be globally zero in every iteration to obtain $\delta_x$ and $\delta_y$.
We empirically define the cutoff between weak and bright events to be $\text{S/N}=60$ and verify that the fit is not very sensitive to this threshold.
The results of this procedure are shown in Fig.~\ref{fig:loc_uncert}, and yield
$\delta_x=0.16(4)$\,arcmin,
$\delta_y=0.17(4)$\,arcmin, 
$\alpha=1.1(3)$, 
$\Sigma_x=0.19(11)$\,arcmin and
$\Sigma_y=0.19(14)$\,arcmin.
The value of the reduced chi-square test for the whole pulsar sample after the correction is $\chi^2_\text{red}=1.1$, validating the results of the procedure.
No systematic deviation is observed as a function of DM or X and Y coordinates.
We apply these corrections to the position of FRBs and other sources localized with the baseband data of CHIME/FRB.


\begin{deluxetable}{lccl}
\tablecaption{
Pulses used as a test sample to verify the localization capability of CHIME/FRB. The source name, the number of pulses used for each source, the DM \citep{man05} and the reference for the source position used are reported.
\label{tab:pulsar_sample}}
\tablewidth{0pt}
\tablehead{
\colhead{Source} & \colhead{Pulses} & \colhead{DM} & \colhead{Reference}\\
 & & pc\,cm$^{-3}$ & 
}
\startdata
PSR B0148$-$06 & 2 & 25.7 & \citet{del19} \\
PSR B0329$+$54 & 2 & 26.8 & \citet{del19} \\
PSR B0355$+$54 & 8 & 57.1 & \citet{li16} \\
PSR B0531$+$21 & 3 & 56.8 & \citet{mcn71} \\
PSR B0643$+$80 & 6 & 33.3 & \citet{hob04} \\
PSR B1541$+$09 & 2 & 35.0 & \citet{cha09} \\
PSR B1642$-$03 & 1 & 35.8 & \citet{del19} \\
PSR B1811$+$40 & 5 & 40.8 & \citet{hob04} \\
PSR B2210$+$29 & 3 & 41.6 & \citet{del19} \\
PSR B2310$+$42 & 6 & 44.2 & \citet{cha09} \\
PSR B2315$+$21 & 5 & 74.5 & \citet{del19} \\
PSR J1652$+$2651 & 1 & 17.3 & \citet{lew04} \\
PSR J2139$+$2242 & 1 & 20.9 & \citet{say97} \\
PSR J2346$-$0609 & 1 & 22.5 & \citet{del19} \\
FRB 20180916B & 10 & 348.8 & \citet{mar20} \\
\enddata
\end{deluxetable}

\section{Conclusions}
\label{sec:conclusions}
The automated pipeline to process data stored by the triggered baseband recording system of CHIME/FRB has been presented.
Baseband data are stored automatically following the detection of an FRB candidate by the CHIME/FRB real-time pipeline.
These data are then processed with a dedicated pipeline run using commercial platforms such as Airflow and Docker to make it easily scalable, portable and making optimal use of machine resources.
The pipeline is able to produce science output without human intervention on a hourly timescale on our machines.

After the initial parameters of the candidate burst are refined, the baseband pipeline localizes its position on the sky.
A grid of tied-beam arrays is formed around the initial position and the resulting intensity map is fitted with a 2D Gaussian model.
Localization of a sample of known sources enabled us to validate the technique and to characterize localization uncertainties.
The best localization precision currently achievable for bright bursts is $\sim 11$\,arcsec.
The results of this localization algorithm are being used for different scientific goals such as correlating FRB positions with large-scale structures in the Universe and localizing low-DM FRBs to host galaxies.
We plan to further reduce the impact of systematic effects in the future by improving our modelling and increasing the test sample of known sources used.

Finally, the pipeline produces a beamformed baseband array of data resolved in time and frequency to enable different analyses.
These include the study of FRB polarization properties, of spectro-temporal structures of the bursts and of propagation effects such as gravitational lensing.
It will also be used in the future to cross-correlate the signal TOA at multiple CHIME/FRB outrigger telescopes under construction to localize sources with tens of milli-arcsecond precision.
This will not only enable host-galaxy identification, and hence redshift determinations, but even determine the source location within the host for the majority of CHIME/FRB detected FRBs.

\acknowledgments
D.M. is a Banting Fellow. 
R.M. recognizes support from the Queen Elizabeth II Graduate Scholarship and the Lachlan Gilchrist Fellowship. 
V.M.K. holds the Lorne Trottier Chair in Astrophysics \& Cosmology, a Distinguished James McGill Professorship and receives support from an NSERC Discovery Grant and Gerhard Herzberg Award, from an R. Howard Webster Foundation Fellowship from CIFAR, and from the FRQNT CRAQ.
C. L. is supported by the U.S. Department of Defense (DoD) through the National Defense Science \& Engineering Graduate (NDSEG) Fellowship.
M.D. is supported by a Killam fellowship, NSERC, CIfAR, and FQRNT.
P.S. is a Dunlap Fellow and an NSERC Postdoctoral Fellow. The Dunlap Institute is funded through an endowment established by the David Dunlap family and the University of Toronto. 
FRB research at UBC is supported by an NSERC Discovery Grant and by the Canadian Institute for Advanced Research. The CHIME/FRB baseband system is funded in part by a CFI John R. Evans Leaders Fund award to IHS. 

\bibliography{sample63}{}

\begin{thebibliography}{}
\expandafter\ifx\csname natexlab\endcsname\relax\def\natexlab#1{#1}\fi
\providecommand{\url}[1]{\href{#1}{#1}}
\providecommand{\dodoi}[1]{doi:~\href{http://doi.org/#1}{\nolinkurl{#1}}}
\providecommand{\doeprint}[1]{\href{http://ascl.net/#1}{\nolinkurl{http://ascl.net/#1}}}
\providecommand{\doarXiv}[1]{\href{https://arxiv.org/abs/#1}{\nolinkurl{https://arxiv.org/abs/#1}}}

\bibitem[{{Bandura} {et~al.}(2016){Bandura}, {Cliche}, {Dobbs}, {Gilbert},
  {Ittah}, {Mena Parra}, \& {Smecher}}]{ban16_fengine}
{Bandura}, K., {Cliche}, J.~F., {Dobbs}, M.~A., {et~al.} 2016, Journal of
  Astronomical Instrumentation, 5, 1641004, \dodoi{10.1142/S225117171641004X}

\bibitem[{{Berger} {et~al.}(2016){Berger}, {Newburgh}, {Amiri}, {Band ura},
  {Cliche}, {Connor}, {Deng}, {Denman}, {Dobbs}, {Fand ino}, {Gilbert}, {Good},
  {Halpern}, {Hanna}, {Hincks}, {Hinshaw}, {H{\"o}fer}, {Johnson}, {Land
  ecker}, {Masui}, {Mena Parra}, {Oppermann}, {Pen}, {Peterson}, {Recnik},
  {Robishaw}, {Shaw}, {Siegel}, {Sigurdson}, {Smith}, {Storer}, {Tretyakov},
  {Van Gassen}, {Vand erlinde}, \& {Wiebe}}]{ber16}
{Berger}, P., {Newburgh}, L.~B., {Amiri}, M., {et~al.} 2016, in Society of
  Photo-Optical Instrumentation Engineers (SPIE) Conference Series, Vol. 9906,
  \procspie, 99060D, \dodoi{10.1117/12.2233782}

\bibitem[{{Chatterjee} {et~al.}(2009){Chatterjee}, {Brisken}, {Vlemmings},
  {Goss}, {Lazio}, {Cordes}, {Thorsett}, {Fomalont}, {Lyne}, \&
  {Kramer}}]{cha09}
{Chatterjee}, S., {Brisken}, W.~F., {Vlemmings}, W.~H.~T., {et~al.} 2009, \apj,
  698, 250, \dodoi{10.1088/0004-637X/698/1/250}

\bibitem[{{CHIME/FRB Collaboration}(2018)}]{chi18_overview}
{CHIME/FRB Collaboration}. 2018, \apj, 863, 48,
  \dodoi{10.3847/1538-4357/aad188}

\bibitem[{{CHIME/FRB Collaboration} {et~al.}(2019{\natexlab{a}}){CHIME/FRB
  Collaboration}, {Amiri}, {Bandura}, {Bhardwaj}, {Boubel}, {Boyce}, {Boyle},
  {. Brar}, {Burhanpurkar}, {Cassanelli}, {Chawla}, {Cliche}, {Cubranic},
  {Deng}, {Denman}, {Dobbs}, {Fandino}, {Fonseca}, {Gaensler}, {Gilbert},
  {Gill}, {Giri}, {Good}, {Halpern}, {Hanna}, {Hill}, {Hinshaw}, {H{\"o}fer},
  {Josephy}, {Kaspi}, {Landecker}, {Lang}, {Lin}, {Masui}, {Mckinven},
  {Mena-Parra}, {Merryfield}, {Michilli}, {Milutinovic}, {Moatti}, {Naidu},
  {Newburgh}, {Ng}, {Patel}, {Pen}, {Pinsonneault-Marotte}, {Pleunis},
  {Rafiei-Ravandi}, {Rahman}, {Ransom}, {Renard}, {Scholz}, {Shaw}, {Siegel},
  {Smith}, {Stairs}, {Tendulkar}, {Tretyakov}, {Vanderlinde}, \&
  {Yadav}}]{chi19_r2}
{CHIME/FRB Collaboration}, {Amiri}, M., {Bandura}, K., {et~al.}
  2019{\natexlab{a}}, \nat, 566, 235, \dodoi{10.1038/s41586-018-0864-x}

\bibitem[{{CHIME/FRB Collaboration} {et~al.}(2019{\natexlab{b}}){CHIME/FRB
  Collaboration}, {Andersen}, {Bandura}, {Bhardwaj}, {Boubel}, {Boyce},
  {Boyle}, {Brar}, {Cassanelli}, {Chawla}, {Cubranic}, {Deng}, {Dobbs},
  {Fandino}, {Fonseca}, {Gaensler}, {Gilbert}, {Giri}, {Good}, {Halpern},
  {Hill}, {Hinshaw}, {H{\"o}fer}, {Josephy}, {Kaspi}, {Kothes}, {Landecker},
  {Lang}, {Li}, {Lin}, {Masui}, {Mena-Parra}, {Merryfield}, {Mckinven},
  {Michilli}, {Milutinovic}, {Naidu}, {Newburgh}, {Ng}, {Patel}, {Pen},
  {Pinsonneault-Marotte}, {Pleunis}, {Rafiei-Ravandi}, {Rahman}, {Ransom},
  {Renard}, {Scholz}, {Siegel}, {Singh}, {Smith}, {Stairs}, {Tendulkar},
  {Tretyakov}, {Vanderlinde}, {Yadav}, \& {Zwaniga}}]{chi19_8repeaters}
{CHIME/FRB Collaboration}, {Andersen}, B.~C., {Bandura}, K., {et~al.}
  2019{\natexlab{b}}, \apjl, 885, L24, \dodoi{10.3847/2041-8213/ab4a80}

\bibitem[{{Cordes} \& {Lazio}(2002)}]{cor02}
{Cordes}, J.~M., \& {Lazio}, T.~J.~W. 2002, ArXiv Astrophysics e-prints

\bibitem[{{Deller} {et~al.}(2019){Deller}, {Goss}, {Brisken}, {Chatterjee},
  {Cordes}, {Janssen}, {Kovalev}, {Lazio}, {Petrov}, {Stappers}, \&
  {Lyne}}]{del19}
{Deller}, A.~T., {Goss}, W.~M., {Brisken}, W.~F., {et~al.} 2019, \apj, 875,
  100, \dodoi{10.3847/1538-4357/ab11c7}

\bibitem[{{Fonseca} {et~al.}(2020){Fonseca}, {Andersen}, {Bhardwaj}, {Chawla},
  {Good}, {Josephy}, {Kaspi}, {Masui}, {Mckinven}, {Michilli}, {Pleunis},
  {Shin}, {Tendulkar}, {Bandura}, {Boyle}, {Brar}, {Cassanelli}, {Cubranic},
  {Dobbs}, {Dong}, {Gaensler}, {Hinshaw}, {Land ecker}, {Leung}, {Li}, {Lin},
  {Mena-Parra}, {Merryfield}, {Naidu}, {Ng}, {Patel}, {Pen}, {Rafiei-Ravandi},
  {Rahman}, {Ransom}, {Scholz}, {Smith}, {Stairs}, {Vanderlinde}, {Yadav}, \&
  {Zwaniga}}]{fon20}
{Fonseca}, E., {Andersen}, B.~C., {Bhardwaj}, M., {et~al.} 2020, \apjl, 891,
  L6, \dodoi{10.3847/2041-8213/ab7208}

\bibitem[{{Hankins}(1971)}]{han71}
{Hankins}, T.~H. 1971, \apj, 169, 487, \dodoi{10.1086/151164}

\bibitem[{{Hessels} {et~al.}(2019){Hessels}, {Spitler}, {Seymour}, {Cordes},
  {Michilli}, {Lynch}, {Gourdji}, {Archibald}, {Bassa}, {Bower}, {Chatterjee},
  {Connor}, {Crawford}, {Deneva}, {Gajjar}, {Kaspi}, {Keimpema}, {Law},
  {Marcote}, {McLaughlin}, {Paragi}, {Petroff}, {Ransom}, {Scholz}, {Stappers},
  \& {Tendulkar}}]{hes18}
{Hessels}, J.~W.~T., {Spitler}, L.~G., {Seymour}, A.~D., {et~al.} 2019, \apjl,
  876, L23, \dodoi{10.3847/2041-8213/ab13ae}

\bibitem[{{Hobbs} {et~al.}(2004){Hobbs}, {Lyne}, {Kramer}, {Martin}, \&
  {Jordan}}]{hob04}
{Hobbs}, G., {Lyne}, A.~G., {Kramer}, M., {Martin}, C.~E., \& {Jordan}, C.
  2004, \mnras, 353, 1311, \dodoi{10.1111/j.1365-2966.2004.08157.x}

\bibitem[{{Kumar} {et~al.}(2019){Kumar}, {Shannon}, {Os{\l}owski}, {Qiu},
  {Bhandari}, {Farah}, {Flynn}, {Kerr}, {Lorimer}, {Macquart}, {Ng},
  {Phillips}, {Price}, \& {Spiewak}}]{kum19}
{Kumar}, P., {Shannon}, R.~M., {Os{\l}owski}, S., {et~al.} 2019, \apjl, 887,
  L30, \dodoi{10.3847/2041-8213/ab5b08}

\bibitem[{{Leung} {et~al.}(2021){Leung}, {Mena-Parra}, {Masui}, {Bandura},
  {Bhardwaj}, {Boyle}, {Brar}, {Bruneault}, {Cassanelli}, {Cubranic},
  {Kaczmarek}, {Kaspi}, {Landecker}, {Michilli}, {Milutinovic}, {Patel},
  {Pleunis}, {Rahman}, {Renard}, {Sanghavi}, {Stairs}, {Scholz}, {Vanderlinde},
  \& {Chime/Frb Collaboration}}]{leu20}
{Leung}, C., {Mena-Parra}, J., {Masui}, K., {et~al.} 2021, \aj, 161, 81,
  \dodoi{10.3847/1538-3881/abd174}

\bibitem[{{Lewandowski} {et~al.}(2004){Lewandowski}, {Wolszczan}, {Feiler},
  {Konacki}, \& {So{\l}tysi{\'n}ski}}]{lew04}
{Lewandowski}, W., {Wolszczan}, A., {Feiler}, G., {Konacki}, M., \&
  {So{\l}tysi{\'n}ski}, T. 2004, \apj, 600, 905, \dodoi{10.1086/379923}

\bibitem[{{Li} {et~al.}(2016){Li}, {Wang}, {Yuan}, {Wang}, {Hobbs}, {Lentati},
  \& {Manchester}}]{li16}
{Li}, L., {Wang}, N., {Yuan}, J.~P., {et~al.} 2016, \mnras, 460, 4011,
  \dodoi{10.1093/mnras/stw1262}

\bibitem[{{Lorimer} {et~al.}(2007){Lorimer}, {Bailes}, {McLaughlin},
  {Narkevic}, \& {Crawford}}]{lor07}
{Lorimer}, D.~R., {Bailes}, M., {McLaughlin}, M.~A., {Narkevic}, D.~J., \&
  {Crawford}, F. 2007, Science, 318, 777, \dodoi{10.1126/science.1147532}

\bibitem[{{Lorimer} \& {Kramer}(2004)}]{lor04}
{Lorimer}, D.~R., \& {Kramer}, M. 2004, {Handbook of Pulsar Astronomy}, Vol.~4
  ({Cambridge University Press})

\bibitem[{{Manchester} {et~al.}(2005){Manchester}, {Hobbs}, {Teoh}, \&
  {Hobbs}}]{man05}
{Manchester}, R.~N., {Hobbs}, G.~B., {Teoh}, A., \& {Hobbs}, M. 2005, \aj, 129,
  1993, \dodoi{10.1086/428488}

\bibitem[{{Manchester} \& {Taylor}(1972)}]{man72}
{Manchester}, R.~N., \& {Taylor}, J.~H. 1972, \aplett, 10, 67

\bibitem[{{Marcote} {et~al.}(2020){Marcote}, {Nimmo}, {Hessels}, {Tendulkar},
  {Bassa}, {Paragi}, {Keimpema}, {Bhardwaj}, {Karuppusamy}, {Kaspi}, {Law},
  {Michilli}, {Aggarwal}, {Andersen}, {Archibald}, {Bandura}, {Bower}, {Boyle},
  {Brar}, {Burke-Spolaor}, {Butler}, {Cassanelli}, {Chawla}, {Demorest},
  {Dobbs}, {Fonseca}, {Giri}, {Good}, {Gourdji}, {Josephy}, {Kirichenko},
  {Kirsten}, {Landecker}, {Lang}, {Lazio}, {Li}, {Lin}, {Linford}, {Masui},
  {Mena-Parra}, {Naidu}, {Ng}, {Patel}, {Pen}, {Pleunis}, {Rafiei-Ravandi},
  {Rahman}, {Renard}, {Scholz}, {Siegel}, {Smith}, {Stairs}, {Vanderlinde}, \&
  {Zwaniga}}]{mar20}
{Marcote}, B., {Nimmo}, K., {Hessels}, J.~W.~T., {et~al.} 2020, \nat, 577, 190,
  \dodoi{10.1038/s41586-019-1866-z}

\bibitem[{{Masui} {et~al.}(2015){Masui}, {Lin}, {Sievers}, {Anderson}, {Chang},
  {Chen}, {Ganguly}, {Jarvis}, {Kuo}, {Li}, {Liao}, {McLaughlin}, {Pen},
  {Peterson}, {Roman}, {Timbie}, {Voytek}, \& {Yadav}}]{mas15}
{Masui}, K., {Lin}, H.-H., {Sievers}, J., {et~al.} 2015, \nat, 528, 523,
  \dodoi{10.1038/nature15769}

\bibitem[{{Masui} {et~al.}(2019){Masui}, {Shaw}, {Ng}, {Smith}, {Vanderlinde},
  \& {Paradise}}]{mas19}
{Masui}, K.~W., {Shaw}, J.~R., {Ng}, C., {et~al.} 2019, \apj, 879, 16,
  \dodoi{10.3847/1538-4357/ab229e}

\bibitem[{{Masui} \& {Sigurdson}(2015)}]{mas15b}
{Masui}, K.~W., \& {Sigurdson}, K. 2015, \prl, 115, 121301,
  \dodoi{10.1103/PhysRevLett.115.121301}

\bibitem[{{McNamara}(1971)}]{mcn71}
{McNamara}, B.~J. 1971, \pasp, 83, 491, \dodoi{10.1086/129160}

\bibitem[{{Newburgh} {et~al.}(2014){Newburgh}, {Addison}, {Amiri}, {Bandura},
  {Bond}, {Connor}, {Cliche}, {Davis}, {Deng}, {Denman}, {Dobbs}, {Fandino},
  {Fong}, {Gibbs}, {Gilbert}, {Griffin}, {Halpern}, {Hanna}, {Hincks},
  {Hinshaw}, {H{\"o}fer}, {Klages}, {Landecker}, {Masui}, {Parra}, {Pen},
  {Peterson}, {Recnik}, {Shaw}, {Sigurdson}, {Sitwell}, {Smecher}, {Smegal},
  {Vand erlinde}, \& {Wiebe}}]{new14}
{Newburgh}, L.~B., {Addison}, G.~E., {Amiri}, M., {et~al.} 2014, in Society of
  Photo-Optical Instrumentation Engineers (SPIE) Conference Series, Vol. 9145,
  \procspie, 91454V, \dodoi{10.1117/12.2056962}

\bibitem[{{Ng} {et~al.}(2017){Ng}, {Vanderlinde}, {Paradise}, {Klages},
  {Masui}, {Smith}, {Bandura}, {Boyle}, {Dobbs}, {Kaspi}, {Renard}, {Shaw},
  {Stairs}, \& {Tretyakov}}]{ng17}
{Ng}, C., {Vanderlinde}, K., {Paradise}, A., {et~al.} 2017, in 2017 XXXIInd
  General Assembly and Scientific Symposium of the International Union of Radio
  Science (URSI GASS), 1--4

\bibitem[{{Petroff} {et~al.}(2015){Petroff}, {Bailes}, {Barr}, {Barsdell},
  {Bhat}, {Bian}, {Burke-Spolaor}, {Caleb}, {Champion}, {Chandra}, {Da Costa},
  {Delvaux}, {Flynn}, {Gehrels}, {Greiner}, {Jameson}, {Johnston}, {Kasliwal},
  {Keane}, {Keller}, {Kocz}, {Kramer}, {Leloudas}, {Malesani}, {Mulchaey},
  {Ng}, {Ofek}, {Perley}, {Possenti}, {Schmidt}, {Shen}, {Stappers}, {Tisserand
  }, {van Straten}, \& {Wolf}}]{pet15}
{Petroff}, E., {Bailes}, M., {Barr}, E.~D., {et~al.} 2015, \mnras, 447, 246,
  \dodoi{10.1093/mnras/stu2419}

\bibitem[{{Petroff} {et~al.}(2016){Petroff}, {Barr}, {Jameson}, {Keane},
  {Bailes}, {Kramer}, {Morello}, {Tabbara}, \& {van Straten}}]{pet16_frbcat}
{Petroff}, E., {Barr}, E.~D., {Jameson}, A., {et~al.} 2016, \pasa, 33, e045,
  \dodoi{10.1017/pasa.2016.35}

\bibitem[{{Platts} {et~al.}(2019){Platts}, {Weltman}, {Walters}, {Tendulkar},
  {Gordin}, \& {Kandhai}}]{pla18}
{Platts}, E., {Weltman}, A., {Walters}, A., {et~al.} 2019, \physrep, 821, 1,
  \dodoi{10.1016/j.physrep.2019.06.003}

\bibitem[{{Rafiei-Ravandi} {et~al.}(2020){Rafiei-Ravandi}, {Smith}, \&
  {Masui}}]{raf20}
{Rafiei-Ravandi}, M., {Smith}, K.~M., \& {Masui}, K.~W. 2020, \prd, 102,
  023528, \dodoi{10.1103/PhysRevD.102.023528}

\bibitem[{{Sayer} {et~al.}(1997){Sayer}, {Nice}, \& {Taylor}}]{say97}
{Sayer}, R.~W., {Nice}, D.~J., \& {Taylor}, J.~H. 1997, \apj, 474, 426,
  \dodoi{10.1086/303446}

\bibitem[{{Spitler} {et~al.}(2016){Spitler}, {Scholz}, {Hessels}, {Bogdanov},
  {Brazier}, {Camilo}, {Chatterjee}, {Cordes}, {Crawford}, {Deneva}, {Ferdman},
  {Freire}, {Kaspi}, {Lazarus}, {Lynch}, {Madsen}, {McLaughlin}, {Patel},
  {Ransom}, {Seymour}, {Stairs}, {Stappers}, {van Leeuwen}, \& {Zhu}}]{spi16}
{Spitler}, L.~G., {Scholz}, P., {Hessels}, J.~W.~T., {et~al.} 2016, \nat, 531,
  202, \dodoi{10.1038/nature17168}

\bibitem[{{Yao} {et~al.}(2017){Yao}, {Manchester}, \& {Wang}}]{yao17}
{Yao}, J.~M., {Manchester}, R.~N., \& {Wang}, N. 2017, \apj, 835, 29,
  \dodoi{10.3847/1538-4357/835/1/29}

\end{thebibliography}
\bibliographystyle{aasjournal}



\end{document}